\documentclass[11pt]{article}

\usepackage{amsmath}
\usepackage{graphicx}
\usepackage{amsfonts}
\usepackage{amssymb}
\usepackage{epsfig}
\usepackage{color}
\usepackage{psfrag}

\newcommand{\EQ}{\begin{equation}}
\newcommand{\EN}{\end{equation}}
\newcommand{\be}{\begin{equation}}
\newcommand{\ee}{\end{equation}}
\newcommand{\bea}{\begin{eqnarray}}
\newcommand{\eea}{\end{eqnarray}}

\newcommand{\var}{\varepsilon}

\begin{document} \setcounter{page}{0} 
\topmargin 0pt
\oddsidemargin 5mm
\renewcommand{\thefootnote}{\arabic{footnote}}
\newpage
\setcounter{page}{0}
\topmargin 0pt
\oddsidemargin 5mm
\renewcommand{\thefootnote}{\arabic{footnote}}
\newpage
\begin{titlepage}
\begin{flushright}
SISSA 17/2007/EP \\
\end{flushright}
\vspace{0.5cm}
\begin{center}
{\large {\bf Particle decay in Ising field theory with magnetic field\footnote{
To appear in the proceedings of the XVth International Congress on Mathematical
Physics, Rio de Janeiro, 6--11 August 2006.}}}\\ 
\vspace{.8cm}
{\large Gesualdo Delfino}\\
\vspace{0.5cm}
{\em International School for Advanced Studies (SISSA),}\\ 
{\em via Beirut 2-4, 34014 Trieste, Italy}\\
{\em Istituto Nazionale di Fisica Nucleare, sezione di Trieste, Italy}\\
\end{center}
\vspace{1.2cm}

\renewcommand{\thefootnote}{\arabic{footnote}}
\setcounter{footnote}{0}

\begin{abstract}
\noindent
The scaling limit of the two-dimensional Ising model in the plane of 
temperature and magnetic field defines a field theory which provides the 
simplest illustration of non-trivial phenomena such as spontaneous symmetry
breaking and confinement. Here we discuss how Ising field theory also gives
the simplest model for particle decay. The decay widths computed in this 
theory provide the obvious test ground for the numerical methods designed to
study unstable particles in quantum field theories discretized on a lattice. 
\end{abstract}
\end{titlepage}

\newpage
\section{Ising field theory}

Quantum field theory provides the natural tool for the characterization of 
universality classes of critical behavior in statistical mechanics. While the
general ideas based on the renormalization group apply to any dimension 
(see e.g. \cite{Cardy}), the two-dimensional case acquired in the last decades
a very special status. Indeed, after the exact description of critical points
was made possible by the solution of conformal field theories \cite{BPZ}, 
it appeared that also specific directions in the scaling region of 
two-dimensional statistical systems can be described exactly \cite{Taniguchi}. 
Additional insight then comes from perturbation theory around these 
integrable directions \cite{nonint}.

The two-dimensional Ising model plays a basic role in the theory of
critical phenomena since when Onsager computed its free energy and 
provided the first exact description of a second order phase transition 
\cite{Onsager}. Its scaling limit in the plane of temperature and magnetic
field defines a field theory -- the Ising field theory -- which provides
the simplest example of non-trivial phenomena such as spontaneous symmetry
breaking and confinement \cite{McW}. Here we will discuss how Ising field 
theory also yields the simplest model for particle decay \cite{decay}.

\vspace{.3cm}
The Ising model is defined on a lattice by the reduced Hamiltonian
\EQ
E=-\frac{1}{T}\sum_{\langle i,j\rangle}\sigma_i\sigma_j-
H\sum_i\sigma_i\,,\hspace{1cm}\sigma_i=\pm 1
\label{lattice}
\EN
so that the partition function is $Z=\sum_{\{\sigma_i\}}e^{-E}$. On a regular 
lattice in more than one dimension, the model undergoes, for a critical 
value $T_c$ of the temperature and for vanishing magnetic field $H$, a second 
order phase transition associated to the spontaneous breakdown of spin 
reversal symmetry. 

In two dimensions the scaling limit of (\ref{lattice}) is
described by the Ising field theory with action
\EQ
{\cal A} = {\cal A}_{CFT}-\tau\int d^2x \, \varepsilon(x)-
h \int d^2x \, \sigma(x)\,.
\label{ift}
\EN
Here ${\cal A}_{CFT}$ is the action of the simplest reflection-positive 
conformal field theory in two dimensions, which corresponds to the Ising 
critical point \cite{BPZ}. The spin operator $\sigma(x)$ with scaling 
dimension $X_\sigma=1/8$ and the energy operator
$\varepsilon(x)$ with scaling dimension $X_\varepsilon=1$ are, together with 
the identity, the only relevant operators present in this conformal theory. 
The couplings $h$ and $\tau$ account for the magnetic field and the deviation
from critical temperature, respectively.

The field theory (\ref{ift}) describes a family of renormalization group 
trajectories flowing out of the critical point located at $h=\tau=0$ (Fig.~1).
Since the coupling conjugated to an operator $\Phi$ has the dimension of a 
mass to the power $2-X_\Phi$, the combination
\EQ
\eta=\frac{\tau}{|h|^{8/15}}
\label{eta}
\EN
is dimensionless and can be used to label the trajectories. In particular,
the low- and high-temperature phases at $h=0$ and the critical isotherm 
$\tau=0$ correspond to $\eta=-\infty,+\infty,0$, respectively.

\begin{figure}
\centerline{
\includegraphics[width=8cm]{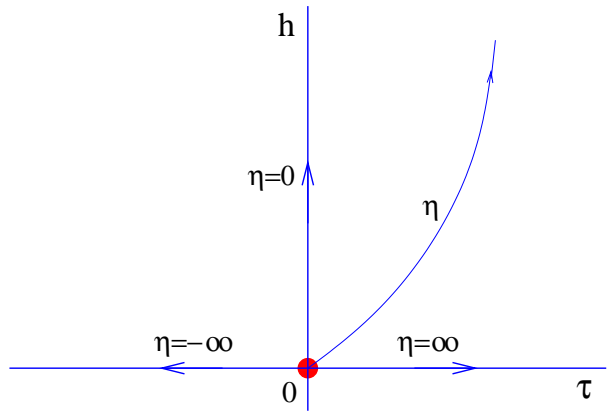}}
\caption{The Ising field theory (\ref{ift}) describes a one-parameter family 
of renormalization group trajectories (labelled by $\eta$) flowing out of the 
critical point located at $\tau=h=0$.}
\end{figure}

\section{Evolution of the mass spectrum}

Under analytic continuation to imaginary time the euclidean field theory
(\ref{ift}) defines a 1+1-dimensional relativistic theory allowing for a 
particle interpretation.

It is well known that the action (\ref{ift}) with $h=0$ describes a free
neutral fermion with mass $m\sim |\tau|$. While in the disordered phase 
$\tau>0$ this fermionic particle corresponds to ordinary spin excitations, 
in the spontaneously broken phase $\tau<0$ it describes the kinks 
interpolating between the two degenerate ground states of the system (Fig.~2a).
In the euclidean interpretation the space-time trajectories of the kinks
correspond to the domain walls separating regions with opposite magnetization.

A small magnetic field switched on at $\tau<0$ breaks explicitely the spin
reversal symmetry and removes the degeneracy of the two ground states 
(Fig.~2b). To first order in $h$ the energy density difference between the two 
vacua is
\EQ
\Delta{\cal E}\simeq 2h\,\langle\sigma\rangle\,,
\EN
where $\langle\sigma\rangle$ is the spontaneous magnetization at $\tau=0$.
With the symmetry broken the kinks are no longer stable excitations. An
antikink-kink pair, which was a two-particle asymptotic state of the theory
at $h=0$, now encloses a region where the system sits on the false vacuum
(Fig.~2b). The need to minimize this region induces an attractive potential
\EQ
V(R)\simeq \Delta{\cal E}\,R
\label{V}
\EN
($R$ is the distance between the walls) which confines the kinks and leaves
in the spectrum of the theory only a string of antikink-kink bound states 
$A_n$, $n=1,2,\ldots$, whose masses 
\EQ
m_n=2m+\frac{(\Delta{\cal E})^{2/3}z_n}{m^{1/3}}\,,\hspace{1cm}h\to 0
\label{m_n}
\EN
are obtained from the Schrodinger equation with the potential (\ref{V}). 
The $z_n$ in (\ref{m_n}) are positive numbers determined
by the zeros of the Airy function, $\mbox{Ai}(-z_n)=0$. This non-relativistic
approximation is exact in the limit $h\to 0$ in which $m_n-2m\to 0$. 
The spectrum
(\ref{m_n}) was first obtained in \cite{McW} from the study of the analytic
structure in momentum space of the spin-spin correlation function for
small magnetic field. Relativistic corrections to (\ref{m_n}) have been
obtained more recently\footnote{See also \cite{FZ3} which appeared after this
talk was given.} in \cite{FZ1}.

\begin{figure}
\centerline{
\includegraphics[width=12cm]{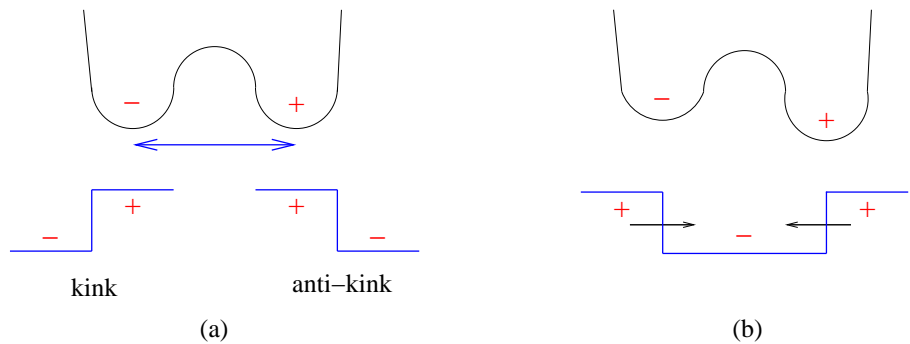}}
\caption{The free energy and the kink excitations in the spontaneously 
broken phase (a). A small magnetic field removes the degeneracy of the ground
states and confines the kinks (b).}
\end{figure}

The particles $A_n$ with mass larger than twice the lightest mass 
$m_1$ are unstable. It was conjectured in \cite{McW}
that the number of stable particles descreases as $\eta$ increases from 
$-\infty$, until only $A_1$ is left in the spectrum of asymptotic particles 
as $\eta\to +\infty$. The particle $A_1$ would then be the free fermion of the 
theory at $\eta=+\infty$.
According to this scenario, for any $n>1$ there should exist
a value $\eta_n$ for which $m_n$ crosses the decay threshold $2m_1$, so that 
the particle $A_n$ becomes unstable for $\eta>\eta_n$. The natural expectation
is that the values $\eta_n$ decrease as $n$ increases, in such a way that the 
$\tau$-$h$
plane is divided into sectors with a different number of stable particles 
as qualitatively shown in Fig.~3. The trajectories corresponding to the values 
$\eta_n$ are expected to densely fill the plane in the limit $\eta\to-\infty$.

This pattern has been confirmed by numerical investigations of the 
spectrum of the field theory (\ref{ift}) for all values of $\eta$  
\cite{nonint,FZ1,FZ3}. For $\eta\to
-\infty$ the particles $A_n$ with $n$ large can be studied within the 
semiclassical approximation and their decay widths have been obtained
in \cite{Rutkevich,FZ3}.

\begin{figure}
\centerline{
\includegraphics[width=9cm]{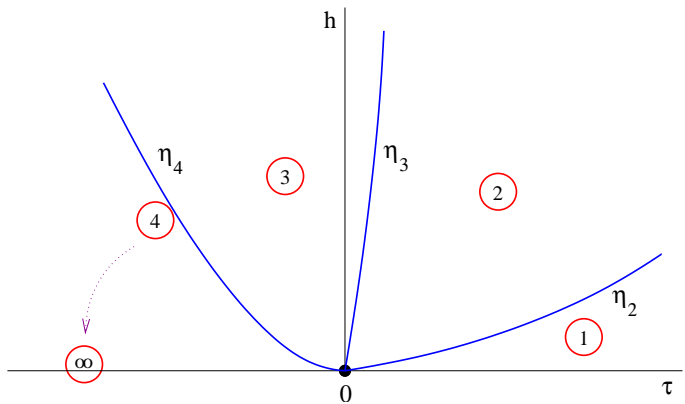}}
\caption{Expected evolution of the mass spectrum as a function of $\eta$. 
In the sector in between $\eta_n$ and $\eta_{n+1}$ the theory possesses $n$ 
stable particles (numbers in the circles).}
\end{figure}

\section{Particle decay off the critical isotherm}

The critical isotherm $\eta=\tau=0$ must lie within the sector in which the 
theory has, generically, three stable particles (Fig.~3). To understand this
we must recall that A.~Zamolodchikov showed in \cite{Taniguchi} that 
the theory (\ref{ift}) with $\tau=0$ is integrable and computed its exact
$S$-matrix. He found that the spectrum along this integrable trajectory 
consists of eight stable particles $A_1,\ldots,A_8$ with 
masses
\bea
m_1 &\sim & h^{8/15} \nonumber \\
m_2 &=& 2 m_1 \cos\frac{\pi}{5} = (1.6180339887..) \,m_1\nonumber\\
m_3 &=& 2 m_1 \cos\frac{\pi}{30} = (1.9890437907..) \,m_1\nonumber\\
m_4 &=& 2 m_2 \cos\frac{7\pi}{30} = (2.4048671724..) \,m_1\nonumber \\
m_5 &=& 2 m_2 \cos\frac{2\pi}{15} = (2.9562952015..) \,m_1\nonumber\\
m_6 &=& 2 m_2 \cos\frac{\pi}{30} = (3.2183404585..) \,m_1\nonumber\\
m_7 &=& 4 m_2 \cos\frac{\pi}{5}\cos\frac{7\pi}{30} = (3.8911568233..) \,m_1\
\nonumber\\
m_8 &=& 4 m_2 \cos\frac{\pi}{5}\cos\frac{2\pi}{15} = (4.7833861168..) \,m_1\,.
\label{e8}
\eea

A peculiarity of this spectrum is that only the lightest three particles lie 
below the lowest decay threshold $2m_1$. The remaining five have the phase 
space
to decay and certainly are not prevented to do so by internal symmetries (the
magnetic field leaves no internal symmetry in the Ising model). It is easy to
see that, while there is nothing wrong with the stability of the particles 
above threshold along this integrable trajectory, they must necessarily decay
as soon as a deviation, however small, from the critical temperature breaks
integrability \cite{decay}. Figure~4 shows the bound state poles and the 
unitarity cuts
of the elastic scattering amplitudes $S_{11}$ and $S_{12}$ in the complex
plane of the relativistic invariant $s$ (square of the center of mass energy).
We know from \cite{Taniguchi} that at $\tau=0$ the scattering channel 
$A_1A_1$ produces the first three particles as bound states (Fig.~4a), while 
the channel $A_1A_2$ produces the first four (Fig.~4b). 
The absence of inelastic scattering in integrable theories allows only for the
unitarity cut associated to the elastic processes. 
When integrability is broken
(i.e. as soon as we move away from $\tau=0$), however, the inelastic channels
and the associated unitarity cuts open up. In particular, the process
$A_1A_2\rightarrow A_1A_1$ acquires a non-zero amplitude, so that the 
threshold located at $s=4m_1^2$ becomes the lowest one also in the $A_1A_2$ 
scattering channel (Fig.~4c). Since the pole associated to $A_4$ is located 
above this threshold, it can no longer remain on the real axis, which in that 
region is 
now occupied by the new cut. The position of the pole must then develop an 
imaginary part which, according to the general requirements for unstable 
particles \cite{ELOP}, is negative and brings the pole through the cut onto 
the unphysical region of the Riemann surface. The other particles above 
threshold, which appear as bound states in other amplitudes at $\tau=0$, 
decay through a similar mechanism. We see then that, beacuse of integrability, 
the trajectory $\eta=0$ corresponds to an isolated case with eight stable 
particles inside a range of values of $\eta$ in which only the particles 
$A_1$, $A_2$, $A_3$ are stable.

These decay processes associated to integrability breaking can be studied 
analytically through the form factor perturbation theory around integrable
models \cite{nonint}. Indeed, if the action of the perturbed integrable
theory is 
\EQ
{\cal A}_{\mbox{integrable}}+\lambda \int d^2x\,\Psi(x)\,,
\label{lambda}
\EN
the perturbative series in $\lambda$ can be expressed in terms of the
matrix elements of the perturbing operator $\Psi$ on the asymptotic particle 
states (Fig.~5). These matrix elements can be computed exactly in the unperturbed, 
integrable theory exploiting analyticity constraints \cite{KW,Smirnov} 
supplemented by operator-dependent asymptotic conditions at high energies 
\cite{immf,DSC,ttbar}. 

\begin{figure}
\centerline{
\includegraphics[width=9cm]{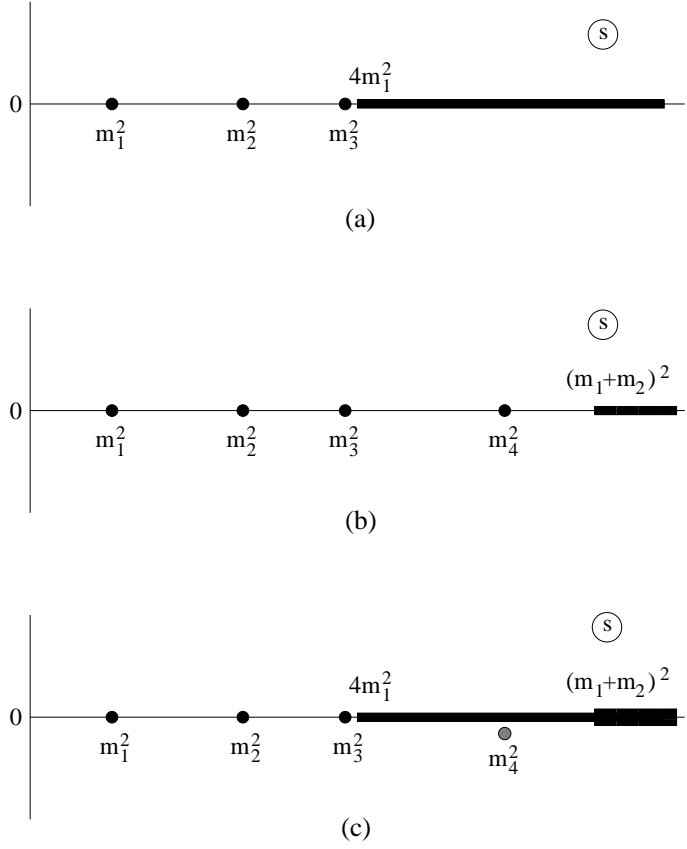}}
\caption{Poles and unitarity cuts for the elastic scattering amplitudes
$S_{11}$ and $S_{12}$ in the integrable case $\tau=0$, (a) and (b), 
respectively, and for $\tau$ slightly different from zero (c). In (c) the 
particle $A_4$ became unstable and the associated pole moved through the cut 
into the unphysical region.} 
\end{figure}

For our present purposes we then look at the action (\ref{ift}) as the 
integrable trajectory $\tau=0$ perturbed by the energy operator 
$\varepsilon(x)$ and must compute corrections in $\tau$. The matrix 
elements of an operator $\Phi(x)$ in the unperturbed theory can all be 
related to the form factors\footnote{The energy and momentum of
the particles are parameterized in terms of rapidities as $(p^0,p^1)=
(m_a\cosh\theta,m_a\sinh\theta)$.}
\EQ
F^\Phi_{a_1\ldots a_n}(\theta_1,\ldots,\theta_n)=
\langle0|\Phi(0)|A_{a_1}(\theta_1)\ldots A_{a_n}(\theta_n)\rangle\,,
\EN
where $|0\rangle$ is the vacuum state and the asymptotic states are built
in terms of the eight particles which are stable at $\tau=0$.

\begin{figure}
\centerline{
\includegraphics[width=11.0cm]{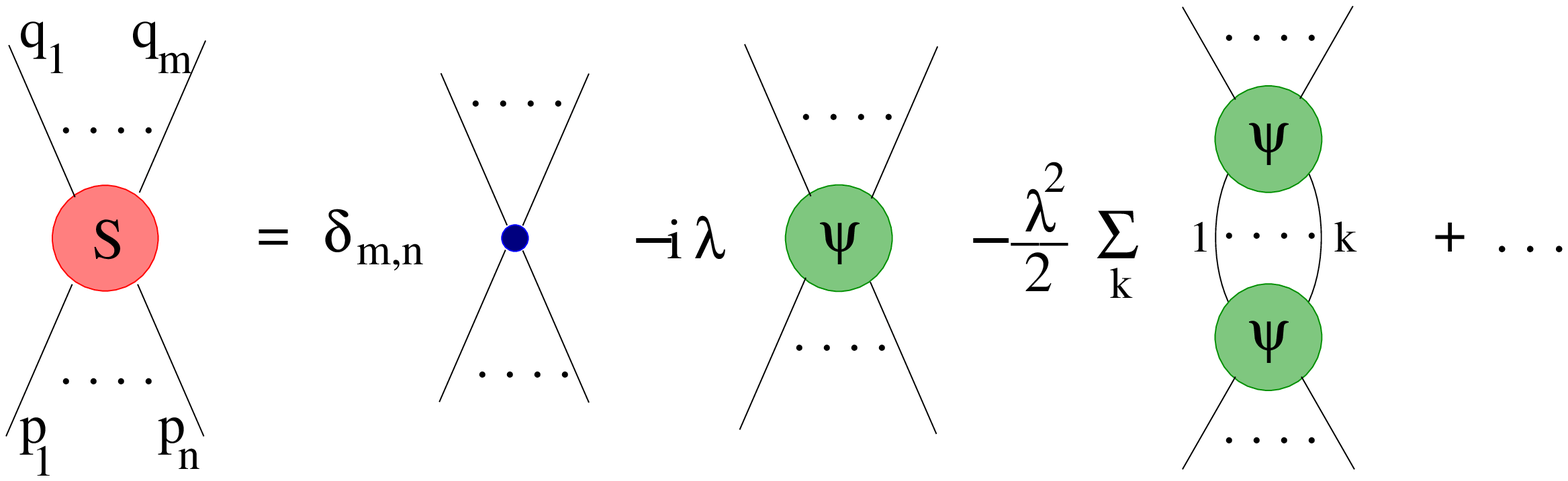}
}
\caption{Perturbative expansion for a scattering amplitude in the theory 
(\ref{lambda}). The matrix elements of the perturbing operator $\Psi$ can be 
computed exactly in the unperturbed, integrable theory.}
\end{figure}

To lowest order in $\tau$ the corrections to the real and imaginary parts
of the masses come from the perturbative terms in Fig.~6 and are given by 
\cite{nonint}
\EQ
\delta\,\mbox{Re}\,m^2_c\simeq -2\tau\,f_{c}\,,
\label{re}
\EN
\EQ
\mbox{Im}\,m_c^2\simeq -
\tau^2\sum_{a\leq b\,,\,m_a+m_b\leq m_c}2^{1-\delta_{ab}}\,\frac{|f_{cab}|^2}
{m_cm_a\left|\sinh\theta^{(cab)}_a\right|}\,,\hspace{1cm}c=4,5
\label{im}
\EN
with
\EQ
f_c=F^\varepsilon_{cc}(i\pi,0)\,,
\EN
\EQ
f_{cab}=F^\varepsilon_{cab}(i\pi,\theta^{(cab)}_a,\theta^{(cab)}_b)\,;
\EN
$\theta^{(cab)}_a$ is determined by energy-momentum conservation 
at the vertices in Fig.~6b.

\begin{figure}
\centerline{
\includegraphics[width=12cm]{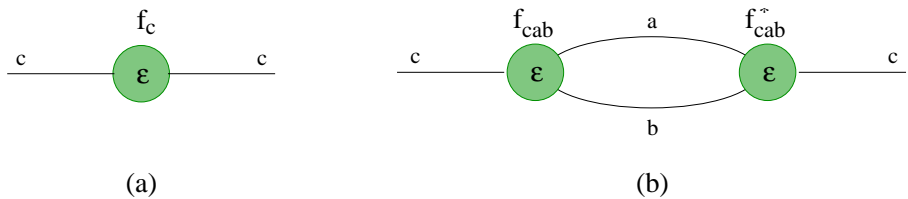}}
\caption{Terms determining the leading corrections to the real (a) and
imaginary (b) parts of the masses in Ising field theory at small $\tau$. For 
$c>5$ also diagrams with more than two particles in the intermediate state
contribute to the imaginary part.}
\end{figure}

The available decay channels for the particles above threshold are determined
by the spectrum (\ref{e8}):
\bea
& & A_4\to A_1A_1\nonumber\\
& &A_5\to A_1A_1,\hspace{.2cm}A_1A_2\nonumber\\
& &A_6\to A_1A_1,\hspace{.2cm} A_1A_2, \hspace{.2cm}A_1A_3,\hspace{.2cm}
 A_1A_1A_1\,,\nonumber
\eea
and similarly for $A_7$ and $A_8$. 
For $c>5$ the sum in (\ref{im}) must be completed including
the contributions of the decay channels with more than two particles in the 
final state.

One- and two-particle form factors for Ising field theory at $\tau=0$ have 
been computed in \cite{immf,DS} (see \cite{review} for a review). 
Table~1 contains the complete list of 
one-particle matrix elements for the relevant operators. The results for 
the lightest particle are compared in Table~2 with the result of numerical 
diagonalization of the transfer matrix on the lattice \cite{CH}. Three-particle
form factors have been computed in \cite{decay} in order to determine the 
imaginary parts (\ref{im}). 

\begin{table}
\begin{center}
\begin{tabular}{|c|r|r|}\hline
  & $\hat{\sigma}$\hspace{1cm} & $\hat{\var}$\hspace{1cm}  \\ \hline\hline
$ F_1 $ & $ -0.640902.. $ & $ -3.706584.. $ \\
$ F_2 $ & $  0.338674.. $ & $  3.422288.. $ \\
$ F_3 $ & $ -0.186628.. $ & $ -2.384334.. $ \\
$ F_4 $ & $  0.142771.. $ & $  2.268406.. $ \\
$ F_5 $ & $  0.060326.. $ & $  1.213383.. $ \\
$ F_6 $ & $ -0.043389.. $ & $ -0.961764.. $ \\
$ F_7 $ & $  0.016425.. $ & $  0.452303.. $ \\
$ F_8 $ & $ -0.003036.. $ &$  -0.105848.. $ \\ \hline
\end{tabular}
\end{center}
\caption{One-particle form factors for the operators $\sigma$ and $\varepsilon$
in Ising field theory at $\tau=0$ \cite{immf,DS}. The rescaling implied by the 
notation 
$\hat{\Phi}\equiv\Phi/\langle\Phi\rangle$ ensures that the results in the table
are universal.}
\end{table}

\begin{table}
\begin{center}
\begin{tabular}{|c||c|c|}\hline
          & Field theory & Lattice \\ \hline
$F_1^{\hat{\sigma}}$ & $-0.640902..$ & $-0.6408(3)$ \\
$F_1^{\hat{\varepsilon}}$ & $-3.70658..$ & $-3.707(7)$ \\
\hline
\end{tabular}
\end{center}
\caption{Lightest-particle form factors at $\tau=0$ from integrable quantum 
field theory \cite{immf,DS} and from numerical diagonalization of the 
transfer matrix \cite{CH}.}
\end{table}

The available results relevant for the leading mass corrections are
\bea
& f_{1}=(-17.8933..)\langle\varepsilon\rangle\hspace{1.5cm}
& |f_{411}|=(36.73044..)\left|\langle\varepsilon\rangle\right|
\nonumber\\
& f_{2}=(-24.9467..)\langle\varepsilon\rangle\hspace{1.5cm}
& |f_{511}|=(19.16275..)\left|\langle\varepsilon\rangle\right|
\nonumber\\
& f_{3}=(-53.6799..)\langle\varepsilon\rangle\hspace{1.5cm}
& |f_{512}|=(11.2183..)\left|\langle\varepsilon\rangle\right|
\nonumber\\
& f_{4}=(-49.3206..)\langle\varepsilon\rangle \hspace{1.5cm}&
\nonumber
\eea
where $\langle\varepsilon\rangle$ is taken at $\tau=0$. The ratios
\EQ
\lim_{\eta\to 0}\frac{\delta\,\mbox{Re}\,m^2_a}{\delta{\cal E}}=2\,\frac{f_a}
{\langle\varepsilon\rangle}\,,
\EN
where $\delta{\cal E}$ is the variation of the vacuum energy density, are 
completely universal and particularly easy to check numerically. These 
predictions for the variations of the real part of the masses of the first
four particles have been confirmed numerically both in the continuum
\cite{nonint,FZ3} and on the lattice \cite{GR}. Using the result\footnote{
The value (\ref{vev}) refers to the normalization of $\varepsilon$ in which
$\langle\varepsilon(x)\varepsilon(0)\rangle\to|x|^{-2}$
as $|x|\rightarrow 0$.} \cite{FLZZ}
\EQ
\langle\varepsilon\rangle=-(2.00314..)\,|h|^{8/15}\,,
\label{vev}
\EN
it easy to check that the the sign of the variations
\EQ
\delta r_a=
-\frac{\tau f_1}{m_1 m_a}\left(r_a^2-\frac{f_a}{f_1}\right)+O(\tau^2)
\label{dr}
\EN
of the mass ratios $r_a=\mbox{Re}\,m_a/m_1$ agrees with the McCoy-Wu scenario
(Fig.~7).

\begin{figure}
\centerline{
\includegraphics[width=10.0cm]{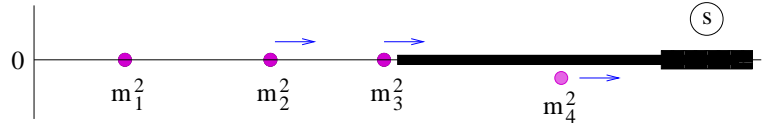}
}
\caption{Evolution of the mass spectrum as predicted by (\ref{dr}) as $\tau$
increases from $0$. $A_2$ and $A_3$ approach the threshold and will decay
for positive values of $\tau$; $A_4$ has already become unstable for $\tau<0$
and moves further away from the threshold.}
\end{figure}

The above values of $f_{abc}$ give \cite{decay}
\EQ
\mbox{Im}\,m_4^2\simeq(-173.747..)\,\tau^2\,,
\EN
\EQ
\mbox{Im}\,m_5^2\simeq(-49.8217..)\,\tau^2\,,
\EN
and in turn the decay widths and lifetimes
\EQ
\Gamma_a=-\frac{\mbox{Im}\,m_a^2}{m_a}\,,\hspace{1.5cm}
t_a=\frac{1}{\Gamma_a}
\EN
of the particles $A_4$ and $A_5$. The lifetime ratio
\EQ
\lim_{\tau\to 0}\frac{t_4}{t_5}\,=0.23326..
\label{liferatio}
\EN
is universal, as well as the branching ratio for $A_5$, which decays at 47\% 
into $A_1A_1$ and for the remaining fraction into $A_1A_2$.

It appears from (\ref{liferatio}) that $A_5$ lives more than
four times longer than $A_4$, somehow in contrast with the expectation 
inherited from accelerator physics that, in absence of symmetry obstructions, 
heavier particles decay more rapidly. Notice, however, that in $d$ dimensions 
the width for the decay $A_c\to A_aA_b$ is
\EQ
\Gamma_{c\to ab}\propto g^2\,|f_{abc}|^2\,\Phi_d\,,
\EN
where $g$ is the perturbative parameter, $f_{abc}$ the form factor and 
\EQ
\Phi_d\sim\int\frac{d^{d-1}\vec{p}_a}{p^0_a}\,\,
\frac{d^{d-1}\vec{p}_b}{p^0_b}\,\,
\delta^d(p_a+p_b-p_c)\sim\frac{|\vec{p}|^{d-3}}{m_c}
\EN
the phase space ($\vec{p}=\vec{p}_a=-\vec{p}_b$). For fixed decay products, 
$|\vec{p}|$ increases with $m_c$ and in $d=2$ suppresses the phase space, in
contrast with what happens in $d=4$. The results for the vertices $f_{abc}$ 
indicate that in our case the dynamics further enhances the increase of 
$t_c$ with $m_c$.

\section{Unstable particles in finite volume}

The issue of obtaining numerical checks of theoretical predictions for 
decay processes is made particularly interesting by the difficulty of 
characterizing unstable particles in the finite volume \cite{Luscher}. The 
problem is particularly relevant for lattice gauge theories.

In two dimensions energy spectra can be obtained by numerical diagonalization
of a truncated Hamiltonian on a cylinder geometry. The signature of  particle 
decay on the cylinder is clear. At an integrable point, when the energy 
levels are plotted as a function of the circumference $R$ of the cylinder, the 
line corresponding to a particle 
above threshold crosses infinitely many levels which belong to the continuum 
when $R=\infty$ (Fig.~8). Once integrability is broken, this line 
``disappears'' through a removal of level crossings and a reshaping of the
lines associated to stable excitations (Fig.~9).

\begin{figure}
\centerline{
\includegraphics[width=11cm]{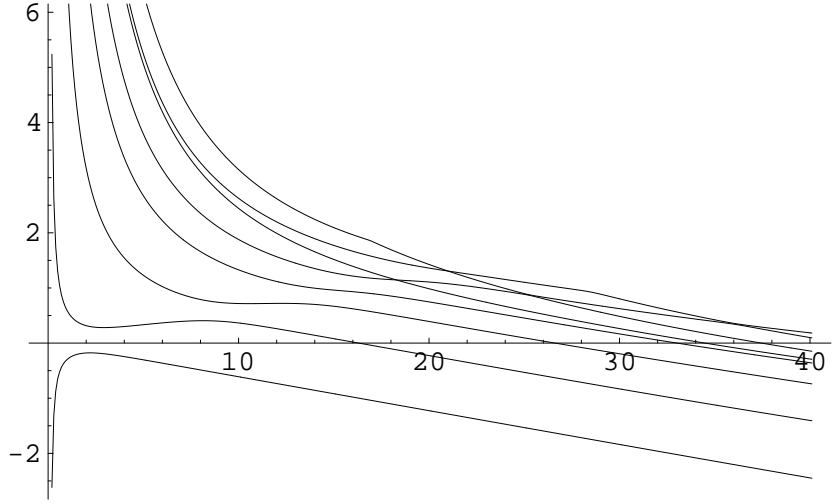}}
\caption{
First eight energy levels of the finite volume Hamiltonian of Ising field
theory at $\tau=0$ as functions of $r =m_1R$ (from Ref.~\cite{nonint}). 
At $r=40$, starting from the bottom, the levels are 
identified as the ground state, the first three particle states $A_1$, $A_2$ 
and $A_3$, three scattering states $A_1A_1$, the particle above threshold 
$A_4$. Crossings between the line associated to the latter and the scattering 
states are visible around $r=18$, $r=25$ and $r=36$.}
\end{figure}

\begin{figure}
\centerline{
\includegraphics[width=11cm]{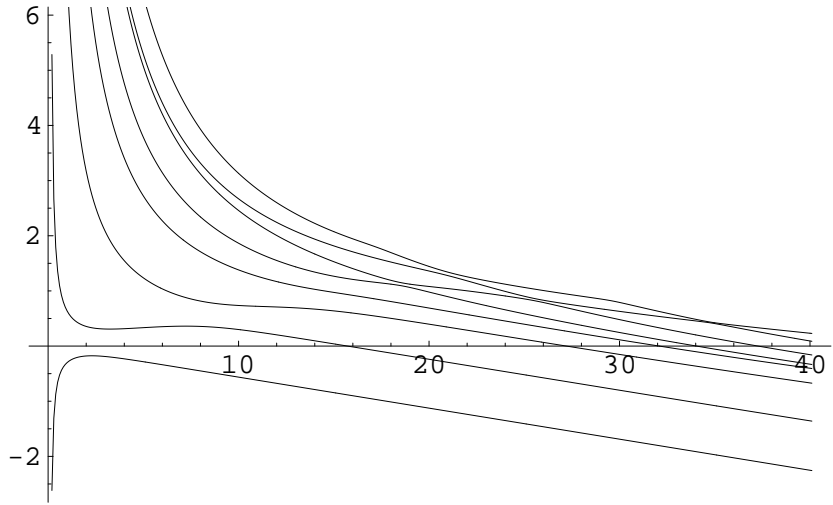}}
\caption{
First eight energy levels of the finite volume Hamiltonian of Ising field
theory slightly away from $\tau=0$ (from Ref.~\cite{nonint}). Observe the 
splitting of the crossings pointed out in the previous figure.}
\end{figure}

One way of extracting the decay width from such a spectrum is the 
following \cite{Luscher}. Consider energy levels corresponding to states 
with two particles of mass $m$ and momenta $p$ and $-p$, sufficiently
close to the threshold $E=2m$ that the particles can only scatter elastically,
with scattering amplitude $S(p)=\exp\,i\delta(p)$.
On the circle, $p$ is quantized by the condition $e^{iRp}\,S(p)=1$, or 
equivalently
\EQ
Rp+\delta(p)=2\pi n\,,
\label{pn}
\EN
with $n$ labelling the states. Hence $\delta(p)$ can be determined from the 
measure of 
\EQ
E(R)=2\sqrt{p^2+m^2}\,.
\EN
A narrow resonance of mass $m_c$ and width $\Gamma$ can then be fitted through
the Breit-Wigner formula
\EQ
\delta(p)=\delta_0(p)+\delta_{BW}(p)\,,\hspace{1.5cm}\delta_{BW}=
-i\ln\frac{E-m_c-i\Gamma/2}{E-m_c+i\Gamma/2}\,,
\EN
$\delta_0$ being a smooth background.

An alternative method was proposed in \cite{decay}. 
At an integrable point consider a particle with mass $m_c>m_a+m_b$. The 
states
\EQ
|A_c(p=0)\rangle\equiv|1\rangle\,,\hspace{1.5cm}
|A_a(p)A_b(-p)\rangle\equiv|2\rangle
\EN
are degenerate on the cylinder at a crossing point $R^*$. 
When integrability is broken by a perturbation 
\EQ
V=\lambda\,\int_0^Rdx\,\Psi(x)\,,
\EN
the energy splitting at $R^*$ is, to lowest order in perturbaton theory,
\EQ
\Delta E=\sqrt{(V_{11}-V_{22})^2+4|V_{12}|^2}\,,
\EN
where $V_{ij}=\langle i|V|j\rangle$.
Choosing a crossing point at $R^*$ large enough, the $V_{ij}$'s are well 
approximated by the infinite volume matrix elements. In this way the decay 
vertices $V_{12}$ can be obtained from measures of the energy splittings.

Both methods have been used in \cite{PT} to measure the decay widths of the
first two particles above threshold in Ising field theory close to the 
critical temperature from numerical diagonalization of a truncated Hamiltonian,
directly in the continuum limit \cite{YZ}. 
The results are shown in Table~3 together with
the predictions of form factor perturbation theory discussed in the previous 
section. It seems obvious that any numerical method designed to measure 
decay widths in lattice models should first of all be able 
to recover these predictions for the two-dimensional Ising model.

\begin{table}
\begin{center}
\begin{tabular}{|c|c|c|}\hline
$|\hat{f}_{cab}|$ & Exact & Numerical  \\ \hline
$ |\hat{f}_{411}| $ & $ 36.730.. $ & $ 36.5(3) $ \\ 
$ |\hat{f}_{511}| $ & $ 19.163.. $ & $ 19.5(9) $ \\ \hline
\end{tabular}
\end{center}
\caption{The values of the three-particle vertices $\hat{f}_{abc}\equiv 
f_{abc}/\langle\varepsilon\rangle$ obtained numerically in \cite{PT} compared 
with the exact predictions of \cite{decay}.}
\end{table}

\vspace{1cm}
{\bf Acknowledgments.}~~I thank P. Grinza and G. Mussardo, my co-authors of
Ref.~\cite{decay} on which this talk is mainly based. This work is partially
supported by the ESF grant INSTANS and by the MIUR project ``Quantum field 
theory and statistical mechanics in low dimensions''.

\end{document}